\documentclass[twocolumn, secnumarabic, amssymb, nobibnotes, prb]{revtex4-1} 

\usepackage{graphicx, amsmath, amssymb}

\newcommand{\super}[1]{$^{\text{#1}}$}	
\newcommand{\sub}[1]{$_{\text{#1}}$}	
\newcommand{\degree}{\ensuremath{^\circ}}	

\begin{document}

\title{Simple method of light-shift suppression in optical pumping systems}%
\author{B. H. McGuyer}%
\author{Y.-Y. Jau}%
\author{W. Happer}%
\affiliation{Department of Physics, Princeton University, Princeton, New Jersey 08544, USA}
\date{\today}%

\begin{abstract}
We report a simple method to suppress the light shift in optical pumping systems.  This method uses only frequency modulation of a radio frequency or microwave source, which is used to excite an atomic resonance, to simultaneously lock the source frequency to the atomic resonance and lock the pumping light frequency to suppress the light shift.  We experimentally validate the method in a vapor-cell atomic clock and verify the results through numerical simulation.  This technique can be applied to many optical pumping systems that experience light shifts.  It is especially useful for atomic frequency standards because it improves long-term performance, reduces the influence of the laser, and requires less equipment than previous methods.
\end{abstract}

\maketitle

The light shift normally refers to the frequency shift of an atomic resonance due to the dynamic (or AC) Stark effect from optical pumping light.\cite{happer:1968} The shift depends on both the frequency and intensity of the light. The light shift is an important source of error in many optical pumping systems. In particular, it is one of the main performance limitations in atomic frequency standards (or clocks) since it turns the fluctuations of the pumping light frequency and intensity into drift and noise in the clock output.  This is one reason why it is difficult to implement atomic clocks with diode lasers. Accordingly, understanding and reducing the light shift in optical pumping systems remains an active area of research.\cite{camparo:2007, vanier:2007, hashimoto:1989, prevwork:signasym}  

There are various ways to suppress light shifts. One simple technique for monochromatic pumping sources, such as diode lasers, is to tune them to a zero-shift optical frequency or ``magic wavelength'' that produces no light shift.\cite{happer:1968}  The zero-shift frequencies are very close to the peaks of the optical absorption lines.  The conventional method to suppress light shifts in laser-pumped atomic clocks uses an additional feedback loop to lock the laser to a zero-shift frequency.\cite{prevwork:cptmod, prevwork:intmod, gong:2008}  In the method we describe here, we use the same feedback loop that locks the local oscillator to the microwave resonance frequency of the atoms to simultaneously adjust the laser frequency to suppress the light shift.    

In conventional atomic clocks, the microwave field is frequency modulated (FM) at a relatively low rate, typically less than 1 kHz.  If the microwave carrier frequency is slightly higher than the atomic resonance frequency, the intensity of the pumping light emerging from the vapor will be modulated at the first harmonic of the FM rate.  If the carrier frequency is slightly lower, the first harmonic component will reverse sign.  Only if the carrier frequency is at exact resonance (the ``zero-crossing frequency'') will there be no first harmonic modulation in the intensity.  The first harmonic component (the signal) is normally detected synchronously with a lock-in amplifier and used as an error signal to lock the microwave carrier frequency (the local oscillator) to the center of the atomic resonance line.

\begin{figure}[b!] 
	\centering
	\includegraphics[]{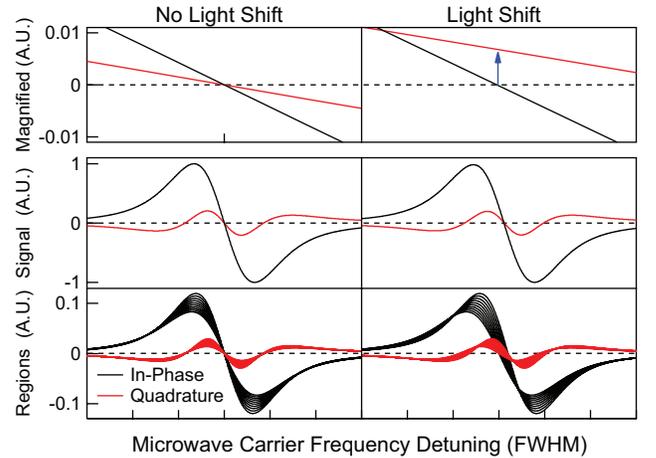}
	\caption{(Color online) Modeling of lock-in channels vs microwave carrier frequency, where the in-phase channel is optimized as the microwave error signal.  An array representing the different regions in a cell with a light-intensity gradient is combined to obtain the total signal.  The top panel shows the zero-crossing region of the middle panel, with both axes magnified.  The arrow indicates the quadrature channel output at the in-phase channel zero-crossing.}
	\label{fig:MWerrorsignals}
\end{figure}

\begin{figure*}[t!] 
	\centering
	\includegraphics[]{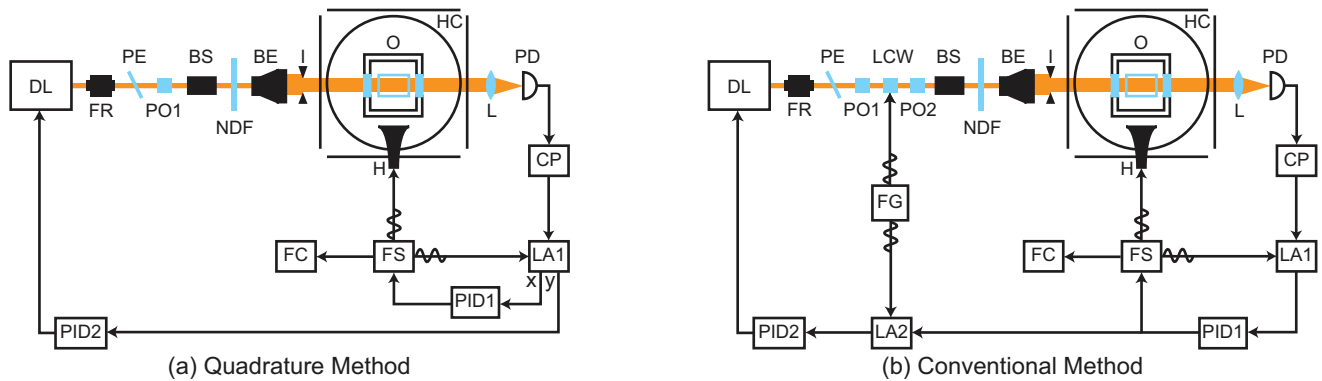}
	\caption{(Color online) Experimental setups with different light-shift suppression methods. DL, diode laser; FR, Faraday rotator; PE, pellicle; PO, polarizer; LCW, liquid crystal wave; BS, beam shaper; NDF, neutral density filter; BE, beam expander; I, iris; O, oven; H, horn; HC, Helmholtz coils; L, lens; PD, photodetector; CP, current preamplifier; LA, lock-in amplifier; PID, PID controller; FS, frequency synthesizer; FC, frequency counter; FG, function generator.}
	\label{fig:expsetup}
\end{figure*}

Superficially, it would seem that the phase adjustment of the lock-in amplifier would not be critical.  However, we were surprised to find that there were often different zero-crossing frequencies for the in-phase and out-of-phase (quadrature) channels of the lock-in amplifier, as shown in Fig.~\ref{fig:MWerrorsignals}.  We eventually found that if the laser was tuned to a zero-shift frequency, the microwave zero-crossing frequency was independent of the lock-in amplifier phase.  This is the phenomenon that is used for the laser stabilization method described in this paper.  We use the in-phase and quadrature channels from the synchronous detector of a single feedback loop as two independent error signals to lock the local oscillator and suppress the light shift.  

The density of atomic vapor used in atomic clocks is normally adjusted to attenuate the laser beam by a factor of about $1/e$.  This means that any light shift will be greater where the beam enters the cell than where it exits.  A spatial beam profile will have a similar effect.  Without a light shift, the entire cell will share a zero-crossing frequency.  However, with a light shift, such light-intensity gradients will lead to regions in the cell with different light shifts and different homogenous broadening, as depicted in Fig.~\ref{fig:MWerrorsignals}.  For stronger light intensities than typical in clocks, this leads to the inhomogenous light shift.\cite{camparo:1983} 

The phase response of the signal from each cell region depends on the linewidth ($\Gamma$), the carrier detuning, and the FM parameters, such as the modulation rate ($\omega_\text{m}$) and index.  When there is a light shift, the different cell regions will not only have different detunings, but will also experience different effective FM rates ($\omega_\text{m} / \Gamma$) due to intensity broadening.  Hence, there is significant phase variation in the signals from each region.  As a result, when there is a light shift, the total signal amplitude will not vanish for any carrier frequency, or in other words, there will be no choice of carrier for which both lock-in channels vanish.  When the in-phase channel is used to lock the local oscillator, the quadrature channel is proportional to the light shift near a zero-shift frequency.  Therefore the quadrature channel may be used as an error signal to lock the pumping light to a zero-shift frequency.  This is the origin of the phenomenon outlined above, which makes it possible to both lock the local oscillator and suppress the light shift with a single feedback loop.  The method is sensitive to the choice of FM parameters and works best if the cell is not optically thin.

A comparison of the ``quadrature'' method with the conventional method to suppress the light shift in a laser-pumped, vapor-cell atomic clock is sketched in Fig.~\ref{fig:expsetup}.  Both methods use two different error signals to lock the local oscillator and to tune the laser to the zero-shift frequency.  Both measure the light-field-independent ground-state hyperfine resonance frequency of \super{87}Rb atoms to about 1 Hz.  The conventional method was recently used to measure small, nonlinear pressure shifts in buffer gases like Ar and Kr that form van der Waals molecules.\cite{gong:2008}

\begin{figure*}[t!]
	\centering
	\includegraphics[]{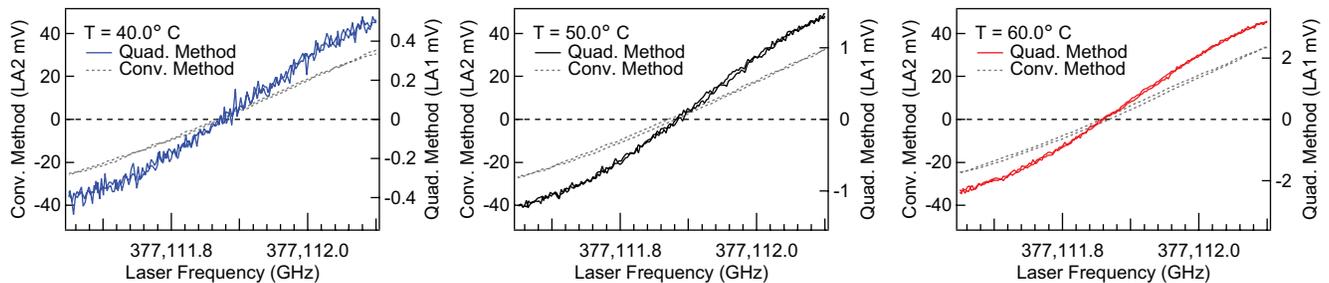}
	\caption{(Color online) Comparison of laser frequency error signals illustrating that the quadrature method locks to the same frequency as the conventional method, which is a zero-shift frequency.  The signal-to-noise ratio of the quadrature method improved at higher temperatures due to increased cell optical thickness.}  
	\label{fig:errorsignalcomparison}
\end{figure*}

Alternating between the two configurations allowed direct comparison between both methods in the same setup.  The experimental setup closely follows the description in Ref.~\onlinecite{gong:2008}. The vapor cell is a cylindrical Pyrex cell, 17 mm in diameter and 25 mm long, filled with a small amount of \super{87}Rb metal.  An external buffer-gas reservoir, pressure gauge, and vacuum port permit convenient changes of the buffer-gas species and pressure.  An air-heated, non-magnetic oven (O) holds the cell at constant temperature between 35--65\degree C.  Helmholtz coils (HC) cancel ambient fields and provide a static longitudinal field of about $0.2$ G. A diode laser (DL) provides 795 nm D1 optical pumping light for \super{87}Rb.  A Faraday rotator (FR) isolates the laser from back-reflected light.  A pellicle (PE) skims off light for analysis with a wavemeter and a Fabry--Per\'{o}t interferometer (not shown).  Polarizers (PO) ensure that the pumping light is linearly polarized.  When included, a liquid crystal wave (LCW) plate driven by a function generator (FG) provides about 30\% intensity modulation of the pumping light at a rate of 2 Hz.  A beam shaper (BS), a beam expander (BE), and an iris (I) ensure that the pumping light fills the cell uniformly.  A rotatable neutral density filter (NDF) adjusts the pumping beam intensity.  A lens (L) collects the transmitted pumping light onto a photodetector (PD).  Microwaves from a frequency synthesizer (FS) are transmitted towards the cell by a horn (H) roughly 10 cm away to drive magnetic resonances. A frequency counter (FC) referenced to a rubidium frequency standard (not shown) measures the microwave frequency.  The microwaves are frequency modulated at a rate of roughly 100--500 Hz with a modulation index of about 1.  A lock-in amplifier (LA1) with a 10--300 ms time constant provides an error signal for a proportional-integral-derivative (PID) controller (PID1) to lock the microwave carrier frequency to the atomic resonance.  

For implementation of the quadrature method (Fig.~\ref{fig:expsetup}(a)), the quadrature channel of LA1 provides an error signal for a PID controller (PID2) to lock the laser to a zero-shift frequency.  For implementation of the conventional method (Fig.~\ref{fig:expsetup}(b)), a second lock-in amplifier (LA2) provides the error signal for PID2 by detecting modulation in the first feedback loop control signal due to intensity modulation from the LCW.  Feedback adjusts the laser frequency through a piezoactuator.  As a quick test for light-shift suppression, we used the NDF filter to temporarily adjust the laser intensity by a factor of 2--4 to verify intensity-independence of the clock output.  

To verify that the quadrature method locks the laser to a zero-shift frequency, we measured the error signals from both methods as a function of laser frequency.  Here, the cell was filled with 30.0 torr of N$_2$ (at 50.0\degree C).  The error signals for both methods were recorded separately with sweeps in both directions, using a wavemeter to record the laser frequency to a precision of 0.01 GHz. Fig.~\ref{fig:errorsignalcomparison} shows the results at 40.0\degree C, 50.0\degree C, and 60.0\degree C, which demonstrate that both methods share the same zero-crossing frequency to within experimental error.  Therefore both methods lock to the zero-shift frequency.  The data also illustrate how higher temperatures improve the signal-to-noise ratio for the quadrature method, which results from increased cell optical thickness.

\begin{figure}[b!] 
	\centering
	\includegraphics[]{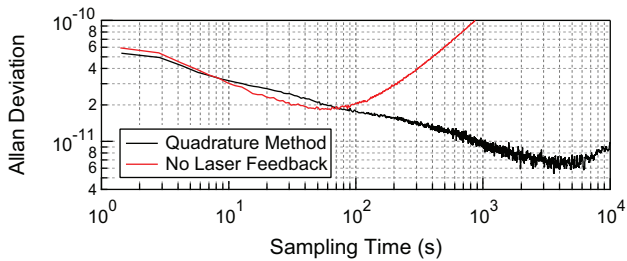}
	\caption{(Color online) Comparison of clock frequency stability with and without the quadrature method of light-shift suppression.}
	\label{fig:avar}
\end{figure}

We measured the clock performance both with and without the quadrature method engaged.  For these tests we used a 27.1 torr buffer-gas mixture of N\sub{2} and Ar at 52.6\degree C, optimized for zero-pressure shift since we were able to control temperature much better than pressure.  Fig.~\ref{fig:avar} shows the Allan deviation for the data.  The clock stability is significantly improved with the quadrature method compared to a free-run system without laser feedback and light-shift suppression. 

\begin{figure}[b!]
	\centering
	\includegraphics[]{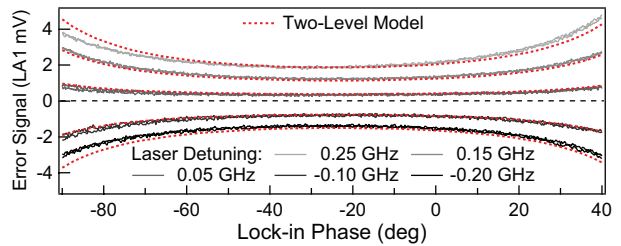}
	\caption{(Color online) Quadrature channel error signals for experiment and modeling versus lock-in phase, with the local oscillator locked.  The different curves denote different fixed laser detuning from the zero-shift frequency.}
	\label{fig:phasescan}
\end{figure}

We modeled this phenomenon in two numerical simulations:  two-level systems and damped simple harmonic oscillators.  Both simulations combined an array representative of regions in a cell with different light shifts and inhomogeneous broadening due to a light-intensity gradient.  This is shown with the two-level model in Fig.~\ref{fig:MWerrorsignals}.  Both simulations agree qualitatively with experimental tests.  In particular, both experiment and modeling show that the selection of lock-in phase is not critical for the quadrature error signal, as shown in Fig.~\ref{fig:phasescan}, where the different curves represent different laser detuning from the zero-shift frequency.  Here, the experimental values are for 30.0 torr N\sub{2} at 50.0\degree C.  Additionally, both experiment and modeling reveal that the quadrature error signal changes sign with increased FM amplitude.  Modeling also indicates that the quadrature error signal is affected by non-light-shift resonance frequency gradients in the cell, such as a temperature gradient.  

While for both methods the response speed of PID2 is limited below that of PID1, in principle this speed may be faster for the quadrature method than for the conventional due to the single modulation scheme.  The quadrature method may also be implemented with the configuration of Fig.~\ref{fig:expsetup}(b) altered to use the quadrature channel of LA1 as the input to LA2.  However, this version sacrifices the original simplicity.  

Although the quadrature method performs well, as shown in Fig.~\ref{fig:avar}, we have observed noticeable shifts of the clock frequency depending on the microwave power and the choice of hyperfine multiplet for optical pumping when the buffer-gas pressure is less than about 20 torr.  The conventional method does not show such dependence.  The clock frequency difference between the two methods at low buffer-gas pressures can be up to a few tens of Hz for 2 torr of Ar or N\sub{2}.  Detailed understanding of these discrepancies requires further investigation.  

We have demonstrated a simple method to suppress the light shift in optical pumping systems, which can be readily applied to existing atomic clocks with few additional components.  The method uses only frequency modulation of a radio frequency or microwave source in order to simultaneously lock the source to an atomic resonance and lock the pumping light to suppress the light shift.  In contrast, conventional stabilization of both sources requires two individual modulation schemes, adding complexity.  In principle, this  technique can also work for coherent population trapping interrogated clocks and other optical pumping systems that experience light shifts.

\vspace{2mm}
The authors are grateful to M.~J.~Souza for making the cell and F.~Gong for contributions to the original apparatus.  This work was supported by the Air Force Office of Scientific Research (Grant FA9550-07-1-0103) and the Department of Defense through the National Defense Science and Engineering Graduate Fellowship (NDSEG) program. 


\end{document}